# EXSeQETIC: Expert System to Support the Implementation of eQETIC Model

R. Rossi and P. N. Mustaro

*Abstract*— The digital educational solutions are increasingly used demanding high quality functionalities. In this sense, standards and models are made available by governments, associations, and researchers being most used in quality control and assessment sessions. The eQETIC model was built according to the approach of continuous process improvement favoring the quality management for development and maintenance of digital educational solutions. This article presents two expert systems to support the implementation of eQETIC model and demonstrates that such systems are able to support users during the model implementation. Developed according to two types of shells (SINTA/UFC and e2gLite/eXpertise2go), the systems were used by a professional who develops these type of solutions and showed positive results regarding the support offered by them in implementing the rules proposed by eQETIC model.

## I. INTRODUÇÃO

AS SOLUÇÕES educacionais digitais são capazes de colaborar com a evolução de diversos setores da sociedade. Isso se dá na educação formal, com os cursos de educação a distância mediados pela Internet, e em treinamentos, que podem ser realizados por meio do *e-learning.* Além dessas modalidades, a aprendizagem pode ser mediada por soluções contemporâneas como o *m-learning* (*mobile learning*) e o *u-learning* (*ubiquitous-learning*).

Assim, a educação digital tem se disseminado de forma significativa no cenário global. Verifica-se, inclusive, que 95% dos professores consideraram que o engajamento dos estudantes aumenta quando se utiliza em tempo de transmissão da instrução algum tipo de tecnologia digital [1].

Com um cenário cada vez mais enfático e promissor quanto à influência das TDIC (Tecnologias Digitais de Informação e Comunicação) na educação, torna-se cada vez mais relevante o desenvolvimento e a distribuição destas soluções com elevada qualidade. Isso é reforçado ao se considerar as estimativas que em 2020 98% dos cursos serão constituídos por soluções híbridas [1].

Desta forma, se verifica por parte de governos, organizações e por parte de universidades e pesquisadores ações que colaboram para abordar a questão da qualidade das soluções educacionais digitais. Tais iniciativas, muitas vezes, estabelecem etapas certificatórias e utilizam padrões para avaliação das soluções.

Há modelos e padrões variados apresentados por associações públicas ou privadas, por governos e por pesquisadores como, por exemplo: 1) IHEP (*Quality on the line: Benchmarks for Success in Internet-Based Distance Education*) [2]; 2) ISO/IEC 19796-1:2005 (*Quality Standard for Learning, Education, and Training*) [3]; e 3) Referenciais de Qualidade para a Educação a Distância do MEC [4]. Contudo estes por vezes são fortemente utilizados em etapas de avaliação destas soluções, em etapas certificatórias, deixando de ser aplicados no planejamento e construção de tais soluções.

No entanto, há uma relação relevante entre a qualidade e a engenharia de processos [5]. Isso porque a qualidade está relacionada aos processos estabelecidos e continuamente aperfeiçoados, que são utilizados para desenvolvimento de produtos e de serviços.

Para as organizações que tratam das soluções educacionais digitais, os processos e procedimentos também são relevantes de forma que possam favorecer o intenso gerenciamento das atividades e, consequentemente, a gestão da qualidade. Desta forma, a abordagem de melhoria contínua de processos torna-se relevante para organizações desta natureza.

Neste âmbito é possível considerar o modelo eQETIC, que trata de qualidade para soluções educacionais digitais [6] e favorece a implementação de processos numa abordagem de melhoria contínua para soluções digitais especificas para a educação online. O modelo eQETIC considera um conjunto de práticas que direciona a implementação dos processos sendo estas distribuídas por três níveis de melhoria: 1) Suficiente, 2) Intermediário e 3) Global.

Embora este modelo possa direcionar a implementação dos processos, muitas vezes as organizações apresentam dificuldades quanto a um diagnóstico prévio sobre a situação atual de seu conjunto de processos. Isso também ocorre com as melhorias em relação aos processos a serem implementados, bem como ao modelo a seguir, e quais seriam os melhores mecanismos para aferir e divulgar a qualidade. Como forma de atender esta necessidade, as organizações buscam por especialistas que podem oferecer serviços e apoiá-las nesta etapa.

Contudo, se verifica que os sistemas especialistas, os quais pertencem ao grupo de sistemas inteligentes baseados em teorias de Inteligência Artificial, podem proporcionar diagnósticos favorecendo o apoio à tomada de decisão [7, 8]. Desta forma, o objetivo desta pesquisa é apresentar dois sistemas especialistas capazes de oferecer suporte a tomada de decisão e implementação do modelo eQETIC. Assim, as organizações podem considerá-lo em seus programas de melhoria de processos.

Para se atender ao objetivo desta pesquisa, é apresentada uma revisão da literatura sobre sistemas especialistas, bem como uma abordagem sobre a relevância da implementação de

R. Rossi, Universidade de São Paulo (USP), São Paulo, São Paulo, Brasil, rossirogerio@hotmail.com
P. N. Mustaro, Universidade Presbiteriana Mackenzie (UPM), São Paulo, São Paulo, Brasil, pollyana.mustaro@mackenzie.br

processos para organizações que desenvolvem soluções educacionais digitais (seção dois); em seguida é detalhada a necessidade de suporte a decisão quanto a utilização do modelo eQETIC para implementação de um programa de melhoria de processos (seção três); a seguir é descrito o experimento realizado (seção quatro); e, ao final, apresenta-se uma discussão sobre os resultados, bem como sugestões sobre possíveis trabalhos futuros (seção cinco).

## II. REVISÃO DA LITERATURA

### A. Sistemas Especialistas

Os Sistemas Especialistas (SE) correspondem a um ramo da Inteligência Artificial (IA) que trata do conhecimento específico para a solução de problemas. O conhecimento neste tipo de sistema é armazenado em bases denominadas bases de conhecimento e favorece a solução de problemas que são apresentadas aos usuários por meio das interfaces do usuário.

Os SE, em geral, atendem a um domínio de conhecimento, pois possuem restrições quanto à formatação de suas bases de conhecimento, que correspondem a uma das maiores dificuldades na implementação deste tipo de sistema.

Segundo [9, p.1], "um sistema especialista é um sistema de computador que simula a capacidade de tomada de decisão de um especialista humano". Isto significa que o sistema deve agir, em todos os aspectos, como o especialista humano.

Os componentes típicos de um SE [9], são apresentados na Fig. 1 e descritos abaixo:

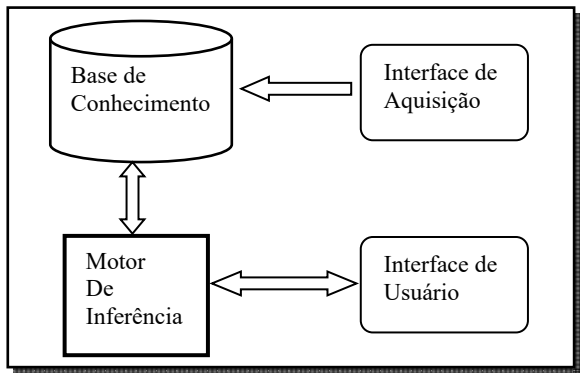

Figura 1. Estrutura básica de um Sistemas Especialista [9].

- **Base de conhecimento**: contempla o domínio do conhecimento necessário para solucionar os problemas apresentados pelo usuário. É codificada em forma de regras de produção, paradigma mais popular para representação do conhecimento. Destaca-se a possibilidade de uso de outros tipos de representação do conhecimento, tais como: redes semânticas, roteiros esquemáticos, quadros e lógica, etc.;
- **Interface do usuário**: mecanismo de comunicação entre o usuário e o sistema;
- **Motor de inferência**: realiza inferências a partir das regras que satisfaçam os fatos ou objetos apresentados, prioriza as regras satisfatórias, e executa a regra com mais alta prioridade.

As três formas que em geral são utilizadas para a implementação de um SE são:

- **Linguagem de programação**: um tradutor de comandos escritos numa sintaxe específica. Uma linguagem de programação para SE também disponibiliza um motor de inferência para executar as sentenças da linguagem. Dependendo da implementação, o motor de inferência pode utilizar o encadeamento para frente (*forward chaining*) ou o encadeamento para trás (*backward chaining*), ou ambos. O PROLOG é um exemplo de linguagem de programação para implementação de sistemas especialistas;
- **Ferramenta**: uma linguagem que pode ser associada a programas utilitários para facilitar o desenvolvimento, correções de erro de codificação e entrega dos programas aplicativos;
- **Shell**: uma ferramenta especialmente concebida para certos tipos de aplicação em que o usuário somente fornece a base de conhecimento. Há diversos *shells* disponíveis para implementação de SE, tais como, ESTA, SINTA/UFC, e2gLite/eXpertise2GO.

As áreas de aplicação dos SE são variadas, sendo que este tipo de sistema pode ser aplicado em medicina, eletrônica, engenharia, geologia, computação, finanças, dentre muitas outras. É possível se verificar, por exemplo, a aplicação deste tipo de sistema para o gerenciamento de recursos naturais [10], bem como para o diagnóstico de doenças do trigo paquistanês [11], e para suportar o tratamento da diabetes [12].

### B. eQETIC Model: características e aplicação

Os processos que se voltam à área de educação podem ser aplicados nas fases de planejamento, desenvolvimento e manutenção das soluções educacionais digitais, sejam elas: o *e-learning*, a educação a distância e os objetos de aprendizagem. Tais soluções digitais para a área educacional exigem processos que devem ser considerados em suas fases de construção e de manutenção. Processos que devem ser implementados e gerenciados para tratar de funções que correspondem, por exemplo, ao suporte, a tutoria, a gestão, a avaliação das soluções educacionais digitais.

Para a educação baseada nas tecnologias digitais existem diversos *frameworks*. Estes possibilitam a implementação de processos que favoreçam o desenvolvimento das soluções digitais, bem como, em alguns casos, procuram estabelecer critérios de avaliação da qualidade de tais soluções. Em geral estes tipos de *frameworks* são utilizados de forma mais enfática em etapas certificatórias das soluções digitais para a educação.

Alguns destes tipos de *framewroks* para a educação digital são: IHEP (*Quality on the line: Benchmarks for Success in Internet-Based Distance Education*) [2]; ISO/IEC 19796-1:2005 (*Quality Standard for Learning, Education, and Training*) [3]; e Referenciais de Qualidade para a Educação a Distância do MEC [4]; e *Standards and Guidelines for Quality Assurance in the European Higher Education Area* [13].

Os *frameworks* voltados à qualidade das soluções educacionais digitais possuem características diversificadas. Estas podem ser tanto estruturais, quanto de aplicação, apresentando mecanismos para tratar a qualidade dos produtos educacionais digitais de maneiras diversas. Estes mecanismos definem práticas que, por vezes, são relacionadas ao controle da qualidade. Por isso são definidos para serem utilizados *a posteriori*, ou seja, após a finalização do produto, sendo instrumentos importantes para etapas avaliativas e certificatórias. Há casos destes modelos que se voltam a abordagem da garantia da qualidade. Tais propostas buscam implementar os processos voltados a estes tipos de soluções educacionais numa visão de melhoria contínua de processos, como é o caso do modelo eQETIC.

O modelo eQETIC (Modelo de Qualidade para Produtos Educacionais baseado nas Tecnologias de Informação e Comunicação) [6] possui uma abordagem voltada a melhoria contínua dos processos, apresentando três Níveis de Melhoria: 1) Suficiente, 2) Intermediário e 3) Global; e seis Entidades Comuns: 1) Didático-Pedagógica, 2) Tecnologia, 3) Gestão, 4) Suporte, 5) Tutoria e 6) Avaliação em sua estrutura, conforme destacado na Fig. 2.

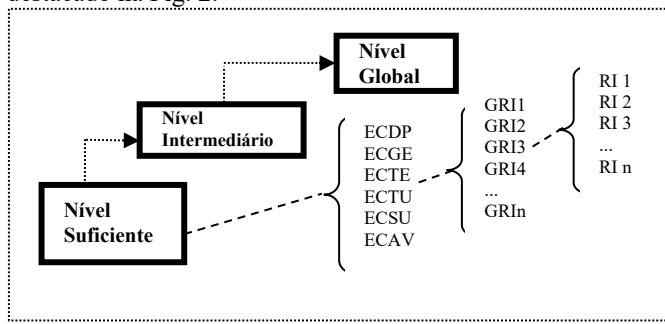

Figura 2. Estrutura do Modelo eQETIC [6].

O modelo eQETIC pode ser considerado por instituições que busquem implementar os processos para o planejamento, desenvolvimento e manutenção de suas soluções educacionais digitais e que concomitantemente busquem construir a qualidade destas.

## III. PROBLEMA DE PESQUISA

A necessidade de uma abordagem sobre a qualidade das soluções educacionais digitais desenvolvidas traz a exigência de normas e padrões para regulamentar a elaboração e a manutenção destes tipos de soluções.

Sendo assim, modelos e padrões são disponibilizados para apoiar as organizações de um modo geral, desde etapas de planejamento e desenvolvimento destas soluções até alcançar a manutenção das mesmas. São estruturas que possuem, por vezes, uma abordagem voltada ao controle de qualidade (por exemplo: IHEP (*Quality on the line: Benchmarks for Success in Internet-Based Distance Education*) e Referenciais de Qualidade para a Educação a Distância do MEC/SEED. Há casos também em que estas voltam-se à garantia da qualidade, permitindo a construção da qualidade durante a elaboração da solução definida pela organização, como por exemplo, o modelo eQETIC.

As organizações que consideram a necessidade de oferecer tais soluções com qualidade devem estabelecer processos para o gerenciamento de suas atividades durante o desenvolvimento e a manutenção destas soluções. Isto implica no estabelecimento prévio de processos específicos que são considerados para o desenvolvimento destas soluções educacionais digitais.

Os processos que envolvem as etapas de planejamento, desenvolvimento e manutenção destes tipos de soluções podem variar. Porém, é possível mapear as atividades de acordo com as seguintes áreas: 1) didático-pedagógicas e psicológicas; 2) processos relacionados ao gerenciamento, sejam estes processos específicos de planejamento ou processos para manter tais soluções; 3) processos relacionados ao gerenciamento da qualidade destas soluções, que abordem questões como tipos de medição e indicadores apropriados que devem ser elaborados e distribuídos; 4) processos relacionados às tutorias que devem ser oferecidas para os usuários deste tipo de solução educacional digital; 5) os processos concernentes ao suporte que deve ser oferecido, bem como processos avaliativos de uma forma geral, sejam numa abordagem formativa ou somativa, referente ao aproveitamento realizado pelo estudante mediante ao curso.

Considerando-se especificamente o Modelo eQETIC, este apresenta um conjunto de regras em sua estrutura, denominadas "Regras de Implementação". Estas são categorizadas em seis entidades, para que sejam utilizadas na implementação dos processos voltados ao desenvolvimento e manutenção das soluções educacionais digitais. Embora todas as regras devam ser implementadas, há uma necessidade de mecanismos de apoio à decisão para as organizações realizarem uma autoavaliação sob possíveis processos que já tenha implementado e que atendam as regras previstas no modelo.

Nota-se que um sistema especialista apresenta tal capacidade, sendo uma ferramenta capaz de amparar este tipo de suporte. Isso porque este tipo de sistema é capaz de operar como um especialista humano em uma dada área do conhecimento [7]. Sendo assim, foi adotada a abordagem de se implementar dois sistemas especialistas (para o estabelecimento de comparação dos resultados) que considerassem, em suas bases de conhecimento, as Regras de Implementação previstas no modelo eQETIC. Com isso buscou-se instituir a possibilidade de oferta de diagnósticos específicos aos seus usuários sobre seus processos em relação ao que é exigido pelo modelo eQETIC.

Para se abordar o experimento realizado e os resultados alcançados, se apresenta na seção seguinte subseções específicas e detalhadas relacionadas a estas atividades.

## IV. EXPERIMENTO E RESULTADOS

Conforme colocado anteriormente, há uma necessidade de suporte à decisão por parte das organizações quanto à implementação de um programa de melhoria de processos. Assim, realizou-se o experimento detalhado a seguir para se verificar a real capacidade de um sistema especialista em apoiar os usuários ao longo do processo de análise e tomada de decisão.

### A. Desenvolvimento dos Sistemas

A partir de estudos sobre os sistemas especialistas, foi possível identificar que estes podem ser implementados por meio de ferramentas do tipo *shell*. Tais estruturas encapsulam a estrutura básica de um sistema especialista (Fig. 1), exigindo somente que se estabeleça a sua base de conhecimento.

No entanto, ao se buscar os *shells* disponíveis para a implementação de sistemas especialistas verificou-se uma série destes, que diferem quanto a alguns pontos, conforme descritos a seguir: 1) interfaces – sejam estas para o engenheiro do conhecimento ou para o usuário do sistema; 2) plataformas que podem ser utilizados; 3) meios de se implementar as bases de conhecimento; e 4) mecanismos do motor de inferência do tipo encadeamento para frente (*forward chaining*) e encadeamento para trás (*backward chaining)*.

Desta forma, o experimento deveria responder as questões que se voltam a: a) capacidade que o sistema especialista, implementado, segundo um determinado *shell*, tem em oferecer um diagnóstico ao usuário e consequente apoio a tomada de decisão; e b) condições oferecidas pelas ferramentas (*shells*) para se implementar com qualidade os sistemas especialistas.

Sendo assim, foram desenvolvidos dois sistemas especialistas baseados nos *shells* SINTA/UFC [14] e e2gLite/eXpertise2go [15] que originaram os seguintes sistemas especialistas:

- EXSeQETIC-SIN – Sistema Especialista para utilização do modelo eQETIC com base no *shell* SINTA/UFC; e
- EXSeQETIC-2GO – Sistema Especialista para utilização do modelo eQETIC com base no *shell* e2gLite/eXpertise2go.

As bases de conhecimentos dos dois sistemas especialistas desenvolvidos, tanto a do EXSeQETIC-SIN quanto a do EXSeQETIC-2GO, consideram as Regras de Implementação da Entidade Comum Didático-Pedagógica, do Nível de Melhoria (Suficiente) definidas pelo modelo eQETIC.

O EXSeQETIC-SIN foi desenvolvido com base no *shell* SINTA/UFC. Um *shell* construído e suportado pelo grupo SINTA (Sistemas Inteligentes Aplicados) do LIA (Laboratório de Inteligência Artificial) da UFC (Universidade Federal do Ceará) [14], para ser aplicado na construção de sistemas especialistas.

Este *shell* apresenta condições para que se implemente bases de conhecimento segundo regras de produção do tipo 'se – então' e considera a forma de utilização do encadeamento para trás (*backward chaining)*. Além disso, permite a utilização de fatores de confiança para as regras de produção. O SINTA/UFC apresenta funcionalidades importantes para o Engenheiro de Conhecimento como mecanismos de depuração da base de conhecimento, bem como a possibilidade de se incluir ajuda *on-line* para cada base.

O EXSeQETIC-SIN é um sistema especialista que apresenta 38 variáveis dos tipos univaloradas e multivaloradas. São 16 variáveis-objetivo definidas para o sistema, sendo estas as variáveis que comandam o sistema e que indicam quais as variáveis serão usadas pelo motor de inferência para fornecer os resultados ou os diagnósticos aos usuários do sistema.

As variáveis que apresentam as perguntas recebem apenas valores determinísticos que formatam as interfaces do sistema. A partir destas variáveis são apresentadas as perguntas por meio das interfaces do sistema e que deverão ser respondidas pelo usuário, permitindo ao sistema um comportamento de um especialista nas regras de implementação do nível suficiente do modelo eQETIC, dado que a partir destas perguntas o sistema poderá oferecer algum diagnóstico para a consulta efetuada.

O EXSeQETIC-SIN considera 47 regras de produção, sendo estas passíveis de variação conforme o conhecimento atribuído ao sistema pelo especialista humano. No caso do EXSeQETIC-SIN, este número sofrerá variações à medida que novas regras de implementação forem acrescentadas ao nível de melhoria suficiente para a entidade comum tratada pelo sistema.

O EXSeQETIC-2GO foi construído com base no *shell* e2gLite/eXpertise2go, ferramenta criada pela eXpertise2GO [15, 16], tendo sido desenvolvida para ser utilizada em plataformas *Web*.

Verificou-se que este *shell* tem sua construção pautada em domínios atuais da tecnologia da computação, dado ter sido construído sob a tecnologia *Java Applet*. Isso permite a criação de bases de conhecimento com acesso por meio de uma interface *Web*, podendo ser utilizada por diversos softwares de navegação como o *Internet Explorer* e *Mozila Firefox*.

O e2gLite/eXpertise2go permite construir bases de conhecimento segundo regras de produção do tipo 'se – então', e o motor de inferência considera a forma de encadeamento das regras de produção do tipo para trás (*backward chaining*). Está arquitetura possibilita a utilização de fatores de confiança por regras de produção.

O sistema especialista EXSeQETIC-2GO, construído com base no e2gLite/eXpertise2go, considera como ponto fundamental o uso do comando '*rule*' onde são definidas as regras de produção no formato '*se – então*'. A utilização do comando '*prompt*' permite a apresentação das questões ao usuário do sistema podendo se receber as respostas por parte do usuário com tipos variados, tais como: *multchoice, forcedchoice, choice* e *allchoice*. A inclusão do comado '*prompt*' é descrita na base de conhecimento após a descrição de todas as regras de produção.

O e2gLite/eXpertise2go define o comando '*goal*' para representar o valor que o motor de inferência deve buscar. Logo, todo sistema construído com base neste *shell* deve descrever ao menos um comando do tipo '*goal*'.

Para o usuário do sistema realizar a consulta ele deve utilizar o botão "iniciar a consulta", devendo responder às questões apresentadas pelo sistema considerando os fatores de certeza. Caso não seja preenchido permitirá ao sistema concluir que o usuário possui 100% de certeza quanto a resposta oferecida (CF = 100% (*certainty factor* = 100%)).

A conclusão do diagnóstico proporcionado pelo sistema especialista EXSeQETIC-2GO ocorre quando o usuário responder às questões apresentadas e definir os fatores de certeza. Em seguida o sistema realiza a inferência e apresenta um parecer sobre a consulta realizada.

A partir da construção dos sistemas, e após exaustivos testes, os mesmos puderam ser utilizados por usuários para verificação dos resultados. Desta forma, elaborou-se um questionário específico com questões distribuídas em duas seções. Para tanto, considerou-se alternativas que correspondem a um determinado grau de concordância por parte do respondente, sendo que a escala de *Likert* [17] foi utilizada como instrumento que definiu as alternativas associadas a cada uma das questões.

### B. Participante

A construção dos dois sistemas especialistas foi realizada para que os mesmos pudessem apoiar o usuário do modelo eQETIC quanto a adoção das práticas previstas no modelo e sua implementação. Seguindo-se os passos relacionados ao desenvolvimento de um sistema especialista conforme [9], realizou-se a atividade de testes e validação da base de conhecimento, sendo que os testes finais favoreceram a disponibilização dos sistemas para serem utilizados pelos usuários.

No entanto, para se realizar a validação dos dois sistemas, optou-se por disponibilizar o mesmo a profissionais especialistas em áreas onde o modelo eQETIC pudesse ser aplicado. Neste escopo, considerou-se instituições educacionais ou empresas que desenvolvem soluções educacionais digitais, como por exemplo o *e-learning* ou que implementam cursos na modalidade a distância para ser utilizado em plataformas digitais.

Após identificar candidatos a usuários dos sistemas especialistas (utilizando estratégias vinculadas à abordagem de estudo de caso [18]), um participante pôde realizar a validação dos sistemas. O profissional em questão integra uma equipe que desenvolve cursos de ensino superior de educação a distância para uma Universidade do setor privado localizada na cidade de São Paulo. Desta forma, foi possível expor ao participante as práticas do modelo eQETIC, seus objetivos e aplicação para que em seguida apresentar os dois sistemas a serem utilizados em etapa de validação quanto a sua operacionalização e funcionalidade.

Os procedimentos realizados para o experimento junto ao participante podem ser verificados em maiores detalhes na subseção seguinte, assim como os resultados alcançados com o experimento são verificados em maiores detalhes na subseção D.

### C. Procedimentos

Os procedimentos realizados referentes à utilização e à validação dos sistemas especialistas EXSeQETIC-SIN e EXSeQETIC-2GO seguiram os seguintes passos:

**1. Abertura da etapa de validação** – a abertura dos trabalhos que se voltam a este experimento foi realizada junto ao participante em seu ambiente de trabalho (desenvolvimento de cursos de educação a distância que são realizados pela Internet). As atividades foram formalizadas para que o profissional participante pudesse dedicar-se integralmente a etapa de validação dos dois sistemas;

**2. Apresentação do modelo eQETIC** – inicialmente foi apresentado ao participante o modelo eQETIC que versa sobre a qualidade das soluções educacionais digitais segundo a abordagem de melhoria contínua de processos. A estrutura do modelo foi apresentada ao profissional participante, assim como sua forma de aplicação e suas práticas que colaboram para a implementação de processos para ambientes em que são desenvolvidas e mantidas soluções educacionais digitais;

**3. Apresentação dos sistemas especialistas** – os dois sistemas especialistas foram apresentados ao participante, bem como foi detalhado junto ao participante os objetivos e funcionalidades destes sistemas e o motivo pelo qual haviam sido elaborados dois sistemas. A interface principal de cada um dos sistemas especialistas pode ser verificada nas Figs. 3 e 4.

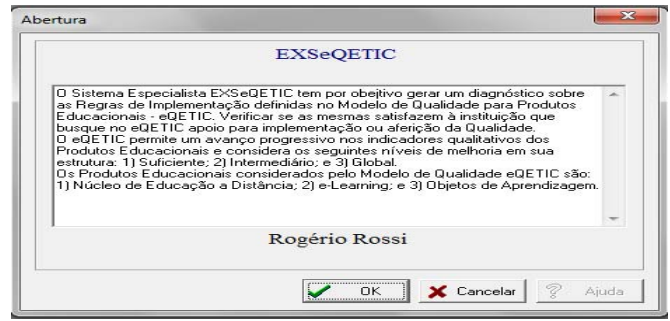

Figura 3. Interface principal do EXSeQETIC-SIN.

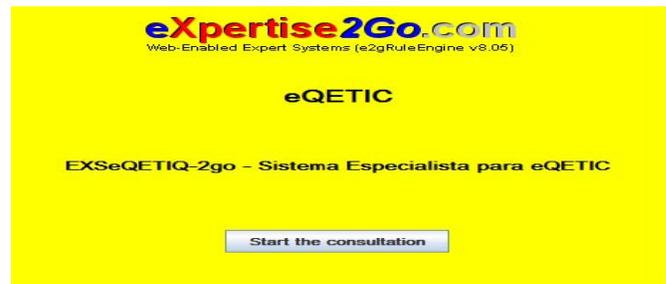

Figura 4. Interface principal do EXSeQETIC-2GO.

**4. Apresentação dos questionários** – os questionários que foram criados para serem utilizados após a utilização dos sistemas foram apresentados ao participante. Em seguida, se explicou ao mesmo que os questionários deveriam ser respondidos após a utilização dos sistemas para se verificar as características de operacionalização e capacidade dos sistemas em oferecer devidamente o diagnóstico;

**5. Utilização dos sistemas especialistas** – os dois sistemas especialistas foram utilizados pelo participante. O uso ocorreu de acordo com as necessidades do participante e os sistemas foram utilizados separadamente, sendo que primeiramente o participante utilizou o sistema EXSeQETIC-SIN e anotou suas considerações quanto ao sistema, realizando em seguida a validação do sistema EXSeQETIC-2GO;

**6. Preenchimento dos questionários** – após terem sido utilizados os dois sistemas o participante foi solicitado a preencher os dois questionários elaborados, um para cada sistema de modo que pudesse formalizar sua validação quanto aos sistemas utilizados;

**7. Encerramento da etapa de validação** – a validação que ocorreu para os dois sistemas especialistas permitiu ao participante comentar suas considerações finais sobre os sistemas e sobre o experimento realizado, apresentando comentários positivos sobre os sistemas desenvolvidos e sobre a eficácia quanto à aplicação dos mesmos.

A realização do experimento seguindo os passos supramencionados permitiu a coleta de dados para sua posterior análise quanto a operacionalização e capacidade que os dois sistemas oferecem em relação a oferta de diagnósticos.

*D. Medição e Análise dos Resultados*

Cabe ressaltar que a coleta realizada envolveu um questionário específico, preenchido após a utilização de cada um dos sistemas especialistas. Os questionários consideravam as mesmas questões, que foram agrupadas para tratar a operacionalidade e usabilidade do sistema na seção I, com seis questões; e, para validar a capacidade do sistema em oferecer suporte à tomada de decisão dez questões foram apresentadas na seção II. Os resultados alcançados que refletem as respostas do participante podem ser identificados na Tabela I.

TABELA I. RESULTADOS DE VALIDAÇÃO DO EXSEQETIC-SIN E DO EXSEQETIC-2GO (POR SEÇÃO).

| Grau de concordância | Sistema Especialista para suporte ao modelo eQETIC | | | |
|---|---|---|---|---|
| | *EXSeQETIC-SIN* | | *EXSeQETIC-2GO* | |
| | *Seção I* | *Seção II* | *Seção I* | *Seção II* |
| CT | 50% | 60% | 16,67 % | 20% |
| CP | 50% | 40% | 50% | 50% |
| NCND | 0% | 0% | 0% | 10% |
| DP | 0% | 0% | 33,33% | 20% |
| DT | 0% | 0% | 0% | 0% |

Legenda: 'CT – Concordo Totalmente'; 'CP – Concordo Parcialmente'; 'NCND – Não Concordo Nem Discordo'; 'DP - Discordo Parcialmente'; 'DT Discordo Totalmente'.

Para as atividades voltadas ao EXSeQETIC-SIN, ou seja, a utilização do sistema pelo participante e suas respostas formalizadas nos questionários quanto ao uso deste sistema, verificou-se que o sistema atendeu às expectativas do participante. Todas as características avaliadas atenderam totalmente ou parcialmente ao participante, conforme se verifica na Tabela I.

Conforme as respostas apresentadas, foi possível se verificar que o participante considerou a operacionalidade e funcionalidade do EXSeQETIC-SIN satisfatória, dados que correspondem ao que se verifica na coluna 'seção I' da Tabela I para o referido sistema. Quanto a capacidade do sistema em oferecer o suporte necessário para a tomada de decisão, o participante também considerou o EXSeQETIC-SIN satisfatório, respondendo a totalidade das questões segundo as escalas 'CT' e 'CP', conforme se verifica na coluna 'seção II' da Tabela I.

Para os resultados que se voltam ao EXSeQETIC-2GO, verificou-se índices de discordância para ambas seções analisadas, conforme é possível se verificar na Tabela I. Tanto para a operacionalidade do sistema quanto para a capacidade em oferecer suporte à tomada de decisão o participante considerou certa discordância, conforme se observa nas colunas 'Seção I' e 'Seção II' para o referido sistema.

Para observarem-se os resultados de uma forma sumarizada quanto a todas as questões respondidas pelo participante, verificam-se na Tabela II as escalas obtidas para cada um dos sistemas.

TABELA II. RESULTADO DE VALIDAÇÃO DO EXSEQETIC-SIN E DO EXSEQETIC-2GO (GERAL).

| Grau de concordância | Sistemas Especialistas para suporte ao modelo eQETIC | |
|---|---|---|
| | *EXSeQETIC-SIN* | *EXSeQETIC-2GO* |
| CT | 56,25% | 18,75% |
| CP | 43,75% | 50,00% |
| NCND | 0,00% | 6,25% |
| DP | 0,00% | 25,00% |
| DT | 0,00% | 0,00% |

Legenda: 'CT – Concordo Totalmente'; 'CP – Concordo Parcialmente'; 'NCND – Não Concordo Nem Discordo'; 'DP - Discordo Parcialmente'; 'DT Discordo Totalmente'.

O sistema EXSeQETIC-SIN, apresenta totalidade de respostas voltadas à concordância total ou parcial quanto a validação e utilização do sistema, enquanto que o sistema EXSeQETIC-2GO apresenta percentuais que se voltam a imparcialidade ou discordância por parte do participante dada a utilização e os testes que realizou para o sistema.

## V. CONCLUSÃO

A utilização dos sistemas especialistas como forma de apoiar o usuário quanto a tomada de decisão sobre as práticas definidas no modelo eQETIC apresentou-se de forma satisfatória segundo os resultados identificados. O participante, além de avaliar a capacidade dos sistemas em oferecer o suporte quanto ao uso do modelo eQETIC, também realizou avaliação quanto operacionalização e funcionalidades do sistema, apresentando parecer favorável também quanto a este quesito para ambos sistemas.

Para o participante, o sistema especialista desenvolvido segundo o *shell* e2gLite, ou seja, o EXSeQETIC-2GO, apresentou-se com menor eficiência. Se verificou certa insatisfação por parte do respondente, apresentando dados que se voltam a discordância parcial do participante quanto ao sistema.

A proposta de desenvolvimento de dois sistemas considerando distintos *shell*s foi satisfatória, pois pôde oferecer visibilidade quanto à utilização destas ferramentas e a relevância que existe na construção dos sistemas especialistas em relação às suas bases de conhecimento.

Funcionalidades e características de cada *shell* são relevantes para o desenvolvimento dos sistemas, dado que por vezes o *shell* oferece recursos técnicos que favorecem a construção do sistema, os testes e a disponibilização para utilização por parte de equipes usuárias.

Dentre os trabalhos futuros, caracteriza-se a construção de sistemas especialistas baseados em outros *shells* e, sobretudo, com bases de conhecimento que considerem todas as Regras de Implementação do modelo eQETIC.

## REFERÊNCIAS

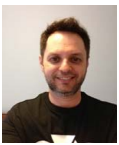
**R. Rossi** é graduado em Matemática pelo Centro Universitário Fundação Santo André (1991), Mestre (1998) e Doutor (2013) em Engenharia Elétrica pela Universidade Presbiteriana. Mackenzie (UPM). Realiza seu Programa de Pós-Doutoramento junto a Escola Politécnica da Universidade de São Paulo (EPUSP) com pesquisas que integram os Sistemas Complexos e Big Data.
Curriculum Lattes: http://lattes.cnpq.br/8288858124711928.

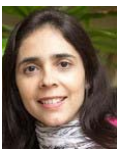
**Pollyana Notargiacomo Mustaro** é graduada (1992), Mestre (1999) e Doutora (2003) em Educação pela USP. Atualmente é Professora da Faculdade de Computação e Informática (FCI) e do Programa de Pós-Graduação em Engenharia Elétrica e Computação (PPGEEC) da Universidade Presbiteriana Mackenzie (UPM).
Curriculum Lattes: http://lattes.cnpq.br/5131975026612008.